\pdfoutput=1
\documentclass[journal]{IEEEtran}
\usepackage{graphicx}
\usepackage{epstopdf}
\usepackage{ifpdf}
\ifpdf
  \DeclareGraphicsExtensions{.pdf,.png,.jpg}
\else
  \DeclareGraphicsExtensions{.eps}
\fi

\graphicspath{{Images/}}
\usepackage{booktabs}
\usepackage{multirow}
\usepackage{todonotes}

\ifCLASSINFOpdf
\else
\fi
\begin{document}
%
\title{Identifying Corresponding Patches in SAR and Optical Images with a Pseudo-Siamese CNN}
%
%
%

\author{Lloyd~H.~Hughes,~\IEEEmembership{Student~Member,~IEEE,}
        Michael~Schmitt,~\IEEEmembership{Senior~Member,~IEEE,}
        Lichao~Mou,
        Yuanyuan~Wang,~\IEEEmembership{Member,~IEEE,}
        and~Xiao~Xiang~Zhu,~\IEEEmembership{Senior~Member,~IEEE}
\thanks{The authors are with Signal Processing
in Earth Observation, Technical University of Munich, 80333 Munich, Germany (e-mail: lloyd.hughes@tum.de, m.schmitt@tum.de, lichao.mou@dlr.de, yuanyuan.wang@dlr.de, xiao.zhu@dlr.de)}
\thanks{X. Zhu is also with the Remote Sensing Technology Institute, German
Aerospace Center, 82234 Wessling, Germany}
\thanks{Manuscript received XXXX YY, ZZZZ.\newline 
This work was supported by the China Scholarship Council, the European Research Council (ERC) under the European Union’s Horizon 2020 research and innovation programme (grant agreement No. ERC-2016-StG-714087, Acronym: \textit{So2Sat}), the Helmholtz Association under the framework of the Young Investigators Group SiPEO (VH-NG-1018, www.sipeo.bgu.tum.de) and the German Research
Foundation (DFG), grant SCHM 3322/1-1.}}

%
%

\markboth{IEEE Geoscience and Remote Sensing Letters, in press}%
{Hughes \MakeLowercase{\textit{et al.}}: Identifying corresponding SAR-optical patches}
%



\maketitle

\begin{abstract}
\textit{This is the pre-print version, to read the final version please go to IEEE Geoscience and Remote Sensing Letters on IEEE Xplore.}\\
In this letter, we propose a pseudo-siamese convolutional neural network (CNN) architecture that enables to solve the task of identifying corresponding patches in very-high-resolution (VHR) optical and synthetic aperture radar (SAR) remote sensing imagery. Using eight convolutional layers each in two parallel network streams, a fully connected layer for the fusion of the features learned in each stream, and a loss function based on binary cross-entropy, we achieve a one-hot indication if two patches correspond or not. 

The network is trained and tested on an automatically generated dataset that is based on a deterministic alignment of SAR and optical imagery via previously reconstructed and subsequently co-registered 3D point clouds. The satellite images, from which the patches comprising our dataset are extracted, show a complex urban scene containing many elevated objects (i.e. buildings), thus providing one of the most difficult experimental environments. The achieved results show that the network is able to predict corresponding patches with high accuracy, thus indicating great potential for further development towards a generalized
multi-sensor key-point matching procedure. 
\end{abstract}


%
\IEEEpeerreviewmaketitle

\section{Introduction}
\begin{figure}[htb]
\centering
\includegraphics[width=0.4\textwidth]{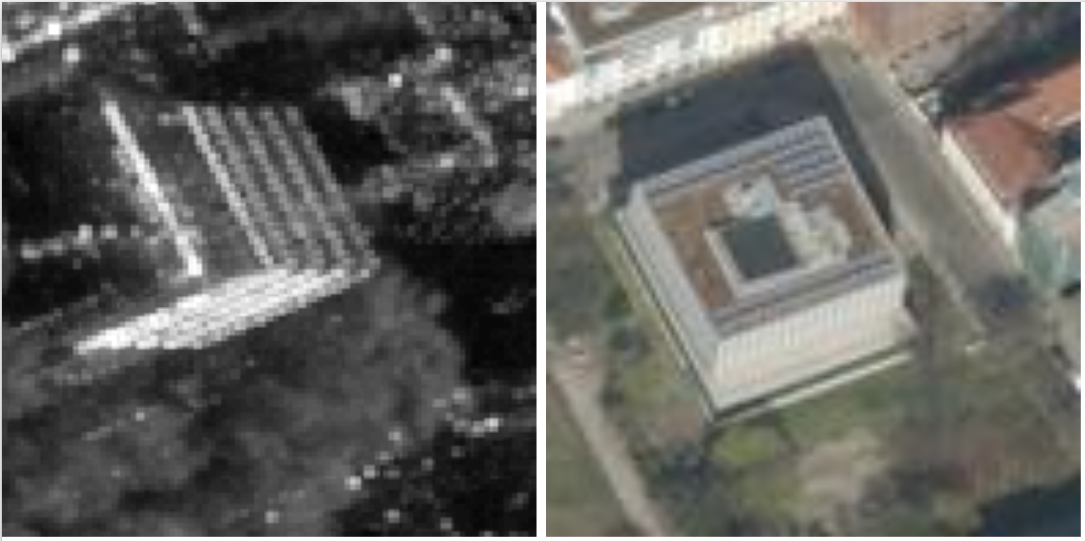}
\caption{A simple detached multi-story building as SAR amplitude image (left) and optical photograph (right).}\label{fig:SAROPT}
\end{figure}

%
%
%
%
\IEEEPARstart{T}{he} identification of corresponding image patches is used
extensively in computer vision and remote sensing-related
image analysis, especially in the framework of stereo applications or co-registration issues. While many successful hand-crafted approaches, specifically designed for the matching of optical
images, exist \cite{Mikolajczyk2005}, to this date the matching of images acquired by
different sensors still remains a widely unsolved challenge \cite{Schmitt2016}. This particularly holds for a joint
exploitation of SAR and optical imagery, caused by
two completely different sensing modalities: SAR imagery
collects information about the physical properties of the scene
and follows a range-based imaging geometry, while optical
imagery reflects the chemical characteristics of the scene and
follows a perspective imaging geometry. Hence,
structures elevated above the ground level, such as buildings or trees, show strongly different appearances in both SAR and optical images (cf. Fig.~\ref{fig:SAROPT}), in particular when dealing with very high resolution data.

In order to deal with the problem of multi-sensor key-point
matching, several sophisticated approaches have been proposed, 
e.g. exploiting phase congruency as a generalization of gradient information \cite{Ye2017}.
However, even sophisticated hand-crafted descriptors reach their limitations for highly resolving data showing densely built-up urban scenes, which -- in the SAR case -- is often difficult to interpret even for trained experts.

Therefore, our work aims at learning a multi-sensor correspondence predictor for SAR and optical image patches of state-of-the-art VHR data. Inspired by promising results achieved in the context of stereo matching for optical imagery \cite{Han2015,Zagoruyko2015}, we also make use of a convolutional
neural network (CNN). The major difference of our work to these purely optical approaches is that we focus on the aforementioned, distinctly
more complicated multi-sensor setup, and therefore design a specific pseudo-siamese network architecture with two separate, yet identical convolutional streams for processing SAR and optical patches
in parallel, instead of a weight-shared siamese network in order
to deal with the heterogeneous nature of the input imagery.

\section{The Network Architecture}

\begin{figure*}[tbh]
\centering
\begin{tabular}{cc}
\includegraphics[width=.65\textwidth]{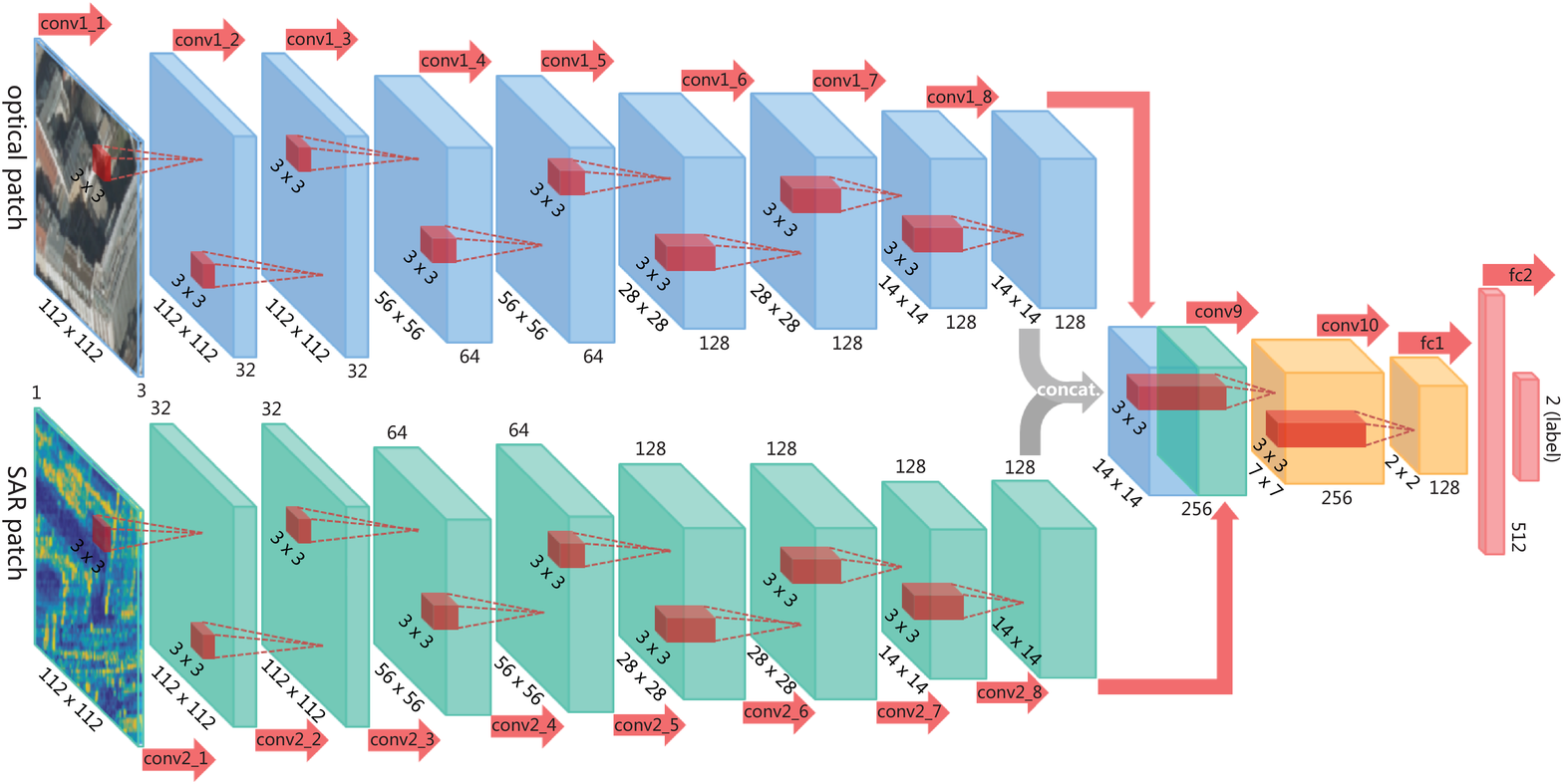} & \raisebox{0.15\height}{\includegraphics[width=.3\textwidth]{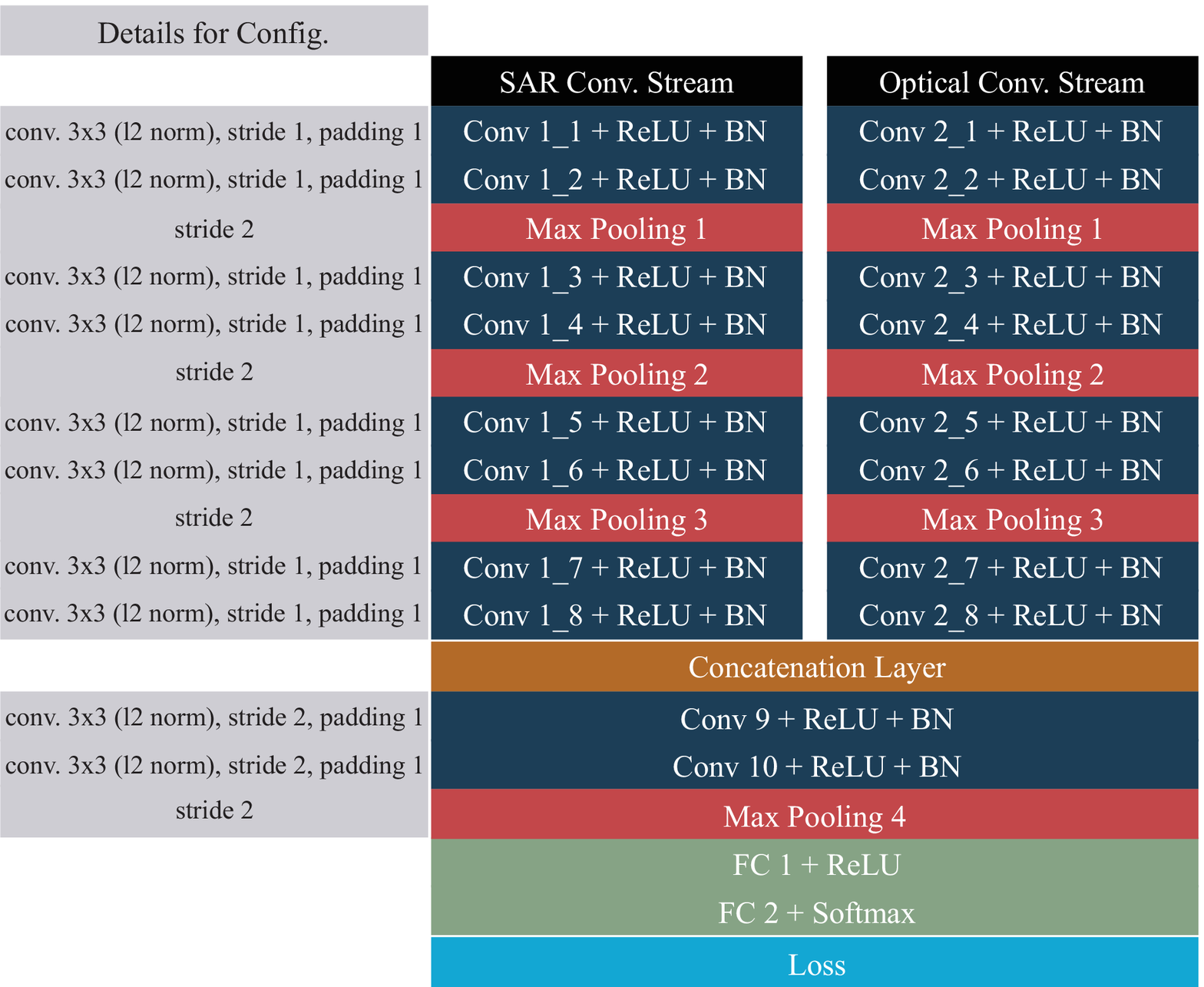}}
\end{tabular}
\caption{Proposed pseudo-Siamese network architecture and layer configuration.}\label{fig:net}
\end{figure*}

\subsection{Pseudo-Siamese Convolutional Network}
Since SAR and optical images lie on different manifolds, it is not advisable to compare them directly by descriptors designed for matching optical patches. Neither suitable are conventional siamese CNN architectures (e.g. \cite{Zagoruyko2015}), which are designed to share their weights in each layer, as the input data processed in the two feature extraction streams share similar properties. In order to cope with the strongly different geometric and radiometric appearances of SAR and optical imagery, in \cite{Mou2017} we proposed a pseudo-siamese network architecture with two separate, yet identical convolutional streams, which process the SAR patch and the optical patch in parallel, and only fuse the resulting information at a later decision stage. Using this architecture, the network is constrained to first learn meaningful representations of the input SAR patch and the optical patch separately, and to combine them on a higher level. The work presented in this letter is an extension of \cite{Mou2017} by improving the fusion part of the network architecture, using a different training strategy, and resorting to non-locally prefiltered SAR patches instead of temporal mean maps. In addition, we evaluate the network on a deterministically partitioned dataset instead of a randomly partitioned one, as random partitioning will always cause positively biased results due to overlapping regions in patches.

The architecture of the proposed network is shown in Fig.~\ref{fig:net}. It is mainly inspired by the philosophy of the well-known VGG Nets \cite{Simonyan2014}. The SAR and optical image patches are passed through a stack of convolutional layers, where we make use of convolutional filters with a very small receptive field of $3\times 3$, rather than using larger ones, such as $5\times 5$ or $7 \times 7$. The reason is that $3\times 3$ convolutional filters are the smallest kernels to capture patterns in different directions, such as center, up/down, and left/right, but still have an advantage: the use of small convolutional filters will increase the nonlinearities inside the network and thus make the network more discriminative

The convolution stride in our network is fixed to 1 pixel; the spatial padding of convolutional layer input is such that the spatial resolution is preserved after convolution, i.e. the padding is 1 pixel for the all $3\times 3$ convolutional layers in our network. Spatial pooling is achieved by carrying out seven max-pooling layers which follow some of the convolutional layers. They are used to reduce the dimensionality of the feature maps. Max-pooling is performed over $2\times 2$ pixel windows with stride 2.

The fusion stage of our proposed network is made up of two consecutive convolutional layers, followed by two fully connected layers. The convolutional layers consist of $3\times 3$ filters which operate over the concatenated feature maps of the SAR and optical streams, in order to learn a fusion rule which minimizes the final loss function. Max-pooling is omitted after the first convolutional layer in the fusion stage, and a stride of 2 is used in order to downsample the feature maps while preserving the spatial information \cite{Handa2016}. The use of $3\times 3$ filters, and the absence of max-pooling after the first convolution allows the fusion layer to learn a fusion rule which is somewhat invariant to spatial mismatches caused by the difference in imaging modalities. This is due to the fact that the fusion layer uses 3x3 convolutions to learn relationships between the features while preserving nearby spatial information. The lack of max pooling means that these learned spatial relationships are preserved as not only the maximal response is considered, while the stride of 2 is used to reduce the feature size. The final stage of the fusion network consists of two fully connected layers: the first of which contains 512 channels; while the second, which performs one-hot binary classification, contains 2 channels.

In a nutshell, the convolutional layers in our network apart from the fusion layer generally consist of $3\times 3$  filters and follow two rules: 1) The layers with same feature map size have the same number of filters; and 2) the number of feature maps increases in the deeper layers, roughly doubling after each max-pooling layer (except for the last convolutional stack in each stream). All layers in the network are equipped with a rectified linear unit (ReLU) as activation function, except the last fully connected layer, which is activated by a softmax function. Figure~\ref{fig:net} shows the schematic diagram of the configuration of our network.

\subsection{Loss Function}
We make use of the binary cross-entropy loss for training our network. Let $X={(x_1^{sar}, x_1^{opt}), (x_2^{sar},x_2^{opt}), ...,(x_n^{sar},x_n^{opt})}$ be a set of SAR-optical patch pairs, where $x_i^{sar},x_i^{opt}\in R^{D\times D},\forall_i=1,..,n$ and $\mathbf{y}_i$ is the one-hot label for the pair $(x^{sar}_i, x^{opt}_i)$ (with $[1, 0]$ denoting a dissimilar pair, and $[0, 1]$ denoting a similar pair). We then seek to minimize the binary cross-entropy loss
\begin{equation}
E=\frac{1}{n}\sum_{i=1}^n\mathbf{y}_i \cdot{\log{f(x^{sar}_i, x^{opt}_i, \theta)}}
\end{equation}
where $f(x^{sar}_i, x^{opt}_i, \theta)$ denotes the output vector of the softmax function when comparing the input pair $(x^{sar}_i, x^{opt}_i)$ with the current network parameters $\theta$.

\subsection{Configuration Details}
Figure \ref{fig:net} depicts the full configuration of our network. Apart from the previously discussed architecture we also make use of batch normalization (BN) after the activation function of each convolutional layer. This leads to an increase in the training speed and reduces the effects of internal covariate shift. In order to reduce over-fitting during training, we made use of $L_2$-regularization, with $\lambda=0.001$, for the convolution kernels of the SAR and optical streams, and dropout with a rate of 0.7 for the first fully connected layer.

\section{Automatic Patch Pool Generation}\label{sec:PatchPool}
For training and testing purposes, a large pool of corresponding and non-corresponding SAR and optical image patches is needed. While the classical work on deep matching for optical imagery can usually rely on easy-to-achieve optical patch pools (see, for example the \textit{Phototourism patch dataset} \cite{Winder2007,Han2015}), annotating corresponding patches in VHR optical and SAR imagery of complex urban scenes is a highly non-trivial task even for experienced human experts. Thus, one of the contributions of this letter is the introduction of a fully automatic procedure for SAR-optical patch pool generation.

\subsection{The ``SARptical'' Framework}
\begin{figure}[htb]
\centering
\includegraphics[width=0.31\textwidth]{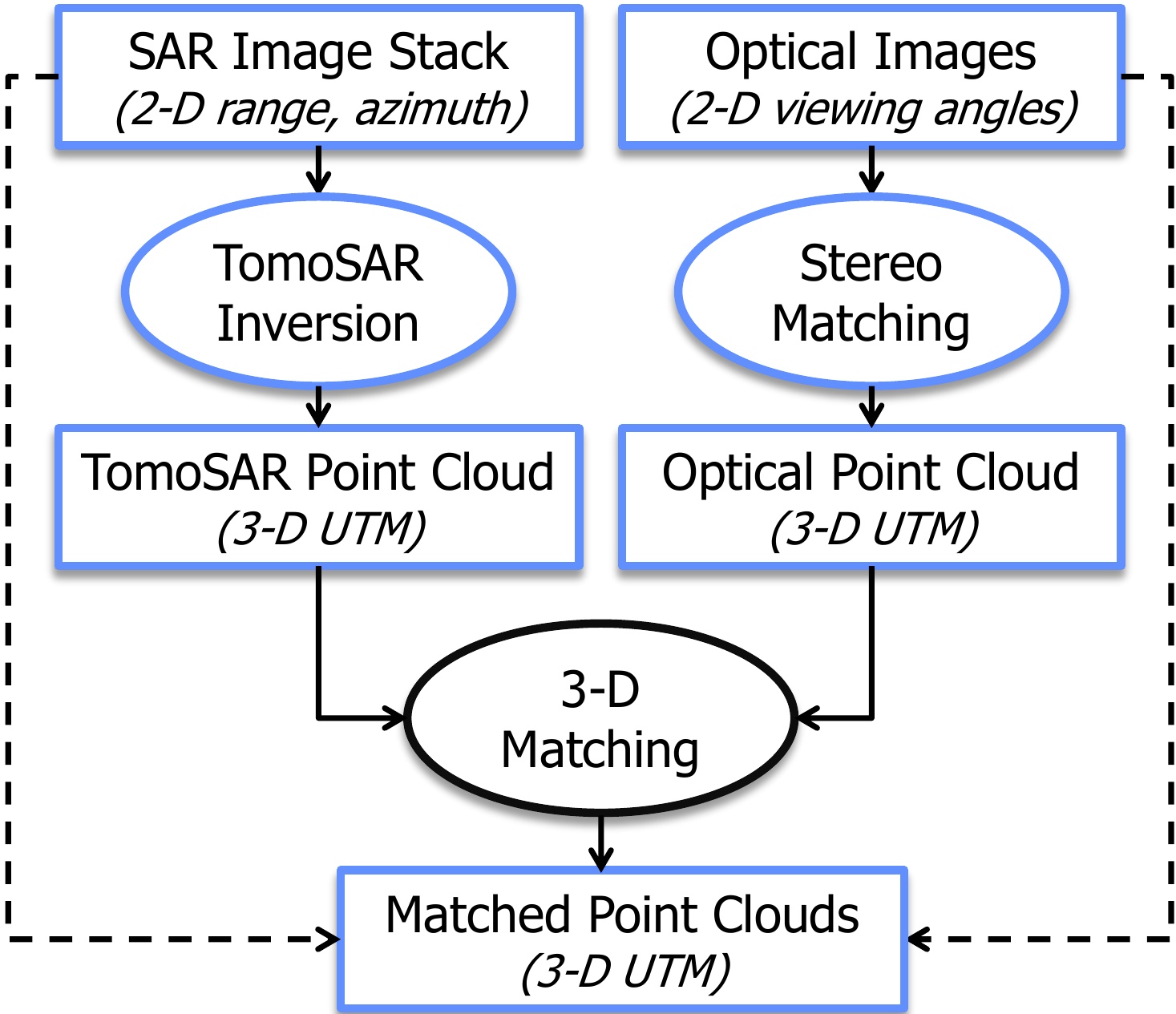}
\caption{Flowchart of the patch-pool generation procedure.}\label{fig:flowchart}
\end{figure}

In order to solve the challenge of automatic dataset generation,
we resort to the so-called ``SARptical'' framework of Wang et al. \cite{Wang2016}, an object-space-based matching procedure
developed for mapping textures from optical images onto 3D
point clouds derived from SAR tomography. The core of
this algorithm is to match the SAR and the optical images in
3D space in order to deal with the inevitable differences caused
by different geometrical distortions. Usually, this would require
an accurate digital surface model (DSM) of the area to link
homologue image parts via a known object space. In contrast,
the approach in \cite{Wang2016} creates two separate 3D point clouds -- one
from SAR tomography and one from optical stereo matching --, which are then registered in 3D space to form a joint (``SARptical'') point cloud, which serves as the necessary representation of
the object space. The flowchart of the approach can be seen in Fig.~\ref{fig:flowchart}.
In order to estimate the 3D positions of the individual
pixels in the images, the algorithm requires an interferometric
stack of SAR images, as well as at least a pair of optical
stereo images. The matching of the two point clouds in 3D
guarantees the matching of the SAR and the optical images.
Finally, we can project the SAR image into the geometry of
the optical image via the ``SARptical'' point cloud, and vice
versa.

\subsection{Data Preparation}
For the work presented in this letter, we made use of a stack of 109 TerraSAR-X
high resolution spotlight images of the city of Berlin, acquired between
2009 and 2013 with about 1 meter resolution, and of 9
UltraCAM optical images of the same area with 20cm ground
spacing. After the reconstruction of the ``SARptical'' 3D point cloud, 8,840 pixels with high SNR ($>5$dB)
were uniformly sampled from the non-locally filtered 
master SAR amplitude image and projected into the individual
optical images, yielding a total of 10,108 corresponding optical pixels. The reason for the difference in pixel numbers is that each of the 9 optical multi-view stereo images is acquired from a different viewing angle, making it possible for each SAR image pixel to have up to 9 corresponding optical image pixels. The actual number of corresponding optical pixels is dependent on the visibility of the SAR pixel from the respective optical point of view. 

All SAR patches are centered at their corresponding SAR image pixels. Their size is fixed at $112 \times 112$ pixels, with a pixel spacing of about 1 m. In analogy, the optical patches are centered at the corresponding optical pixels. After resampling to adjust the pixel spacing, the SAR patches were rotated, so that both patches align with each other as a first approximation. 

In order to reduce bias when training our network we randomly selected a single correct optical correspondence for each SAR image patch during the final dataset preparation. In addition, we randomly assign one wrong optical correspondence to each SAR patch in order to create negative examples. Thus, eventually, we end up with 17,680 SAR-optical patch pairs (see Fig.~\ref{fig:SAROPT} for an example for the class of correct matches).


As final pre-processing steps, the optical patches were converted to gray-scale, 
and all patches were normalized~\cite{WangQi2013} to a radiometric range of $[0;1]$ with subsequent subtraction of their means.

\subsection{Patch Pool Partitioning}
In order to provide a fair experimental design, we partition the patch pool in the following manner: 9724 (55\%) of the patch-pairs are used as training dataset, 2652 (15\%) as validation set, and 5304 (30\%) as test dataset. It has to be noted that we do not partition the patch pool on a purely randomized basis but rather resort to a deterministic partitioning method in order to avoid positively biased test results. The full extent SAR and optical images are first deterministically partitioned and then each partition is processed to generate positive and negative samples for training, validation and testing respectively.

\section{Experiments \& Results}

\subsection{Training Details}
The network was trained using the Adam \cite{Kingma2015} optimization algorithm as it is computationally efficient and exhibits faster convergence than standard stochastic gradient descent methods. The optimization hyper-parameters are fixed to $\beta_1=0.9$, $\beta_2=0.999$ with a learning rate of $\alpha_t=0.0009$. The learning rate was found via a grid search method on our training and validation data, while the $\beta-$parameters were kept at their recommended values. Prior to training the network weight vectors were initialized using the truncated uniform distribution described in \cite{Glorot2010}, and the bias vectors were initialized with zero values. Training was conducted with 2 Nvidia TitanX GPUs using class balanced, mini-batches of 64 SAR-optical patch pairs (32 corresponding, 32 non-corresponding pairs) over 30 epochs; training took on average 25 minutes, with a single forward pass taking around 3ms to complete.

We trained five versions of our proposed network, each at a different patch size, in order to evaluate the effect of patch size on classification accuracy. Patch cropping was done on-the-fly with the new patch being cropped from the center of a larger patch - this was done as the center pixel is the point of correspondence between the SAR and optical patch. Furthermore, we seeded our random number generator with a fixed value of 0, at the start of training for each patch size, in order to prevent the randomization effects between networks.

\subsection{Evaluation Results}\label{sec:EvalResults}
We evaluate the proposed network with different input patch sizes, using our testing patch pool (described in Section \ref{sec:PatchPool}) which has further been cropped around the center pixel to produce new testing pools with different patch sizes.

\begin{figure}[tbh]
\centering
\includegraphics[width=.4\textwidth]{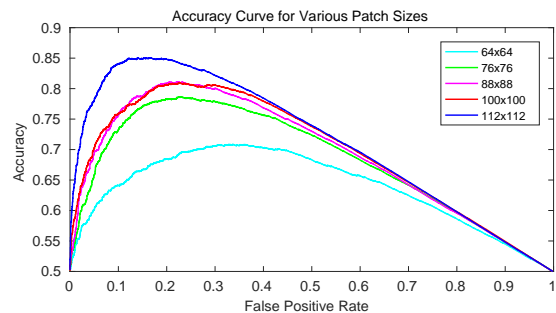}
\caption{Comparison of different patch sizes.}
\label{fig:FPRplot}
\end{figure}

The \textit{accuracy vs. false positive rate} curves corresponding to different patch sizes can be seen in Fig.~\ref{fig:FPRplot}.
Table~\ref{tab:confMatrix} reports the corresponding confusion matrix values for our proposed network evaluated with each patch size; it is to be noted that the confusion matrix is reflective of the network at the point of highest overall performance for each patch size. 

\subsection{Key-Point Matching Results}\label{sec:FeatureResults}
In order to evaluate the proposed network's performance in a real-world, key-point matching scenario we selected 100 neighboring TomoSAR key-points in the SAR image and extracted the corresponding SAR and optical patch pairs. We selected these key-points from a localized area within our test set, so as to reproduce the conditions found in a real-world key-point matching application. We then compared every SAR and optical patch in the selected patch set in order to determine the performance of our proposed network in the presence of large numbers of potential mismatches.

\begin{figure}[htb]
\centering
\begin{tabular}{cc}
\includegraphics[width=0.22\textwidth]{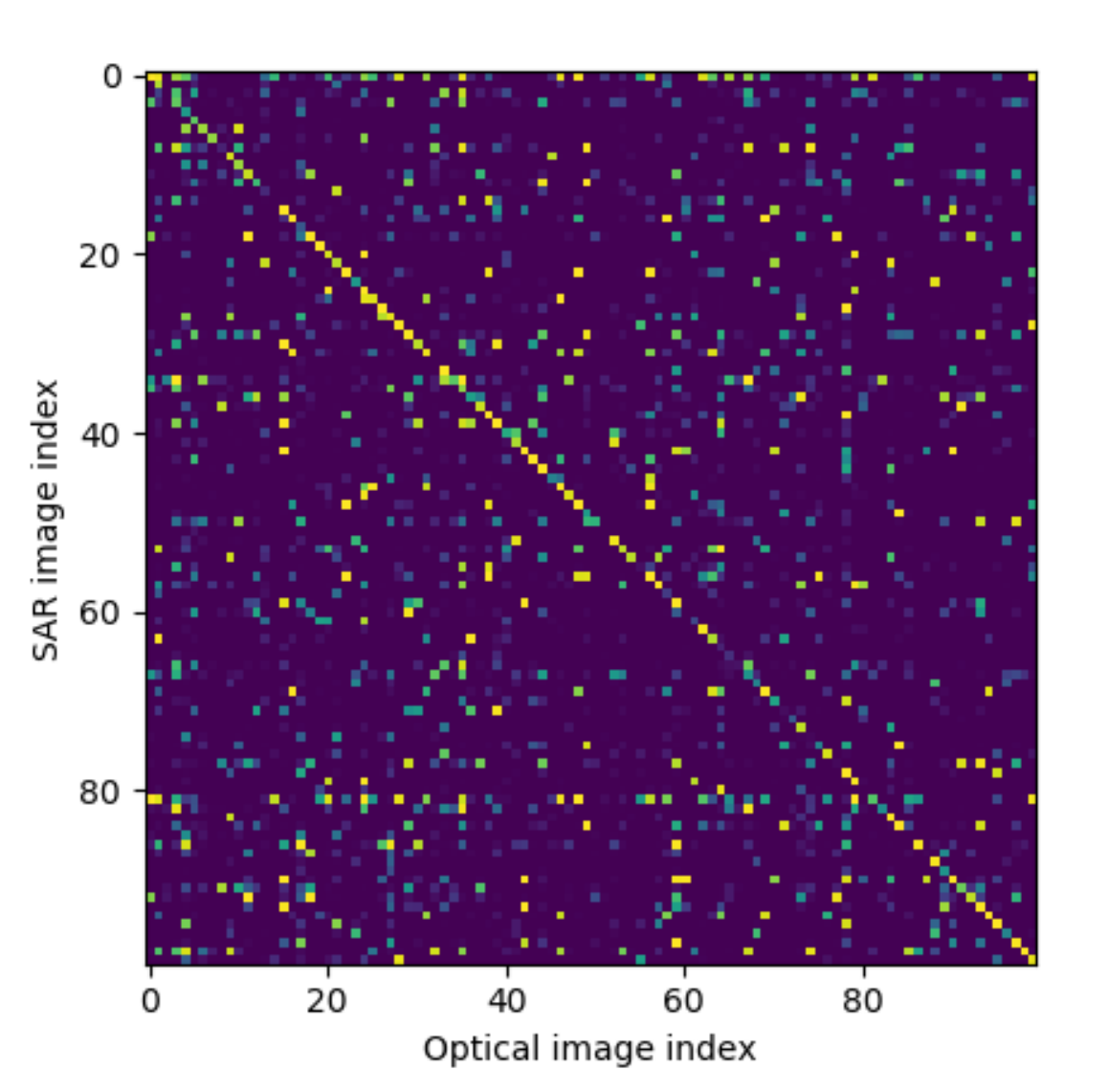} & \includegraphics[width=0.25\textwidth]{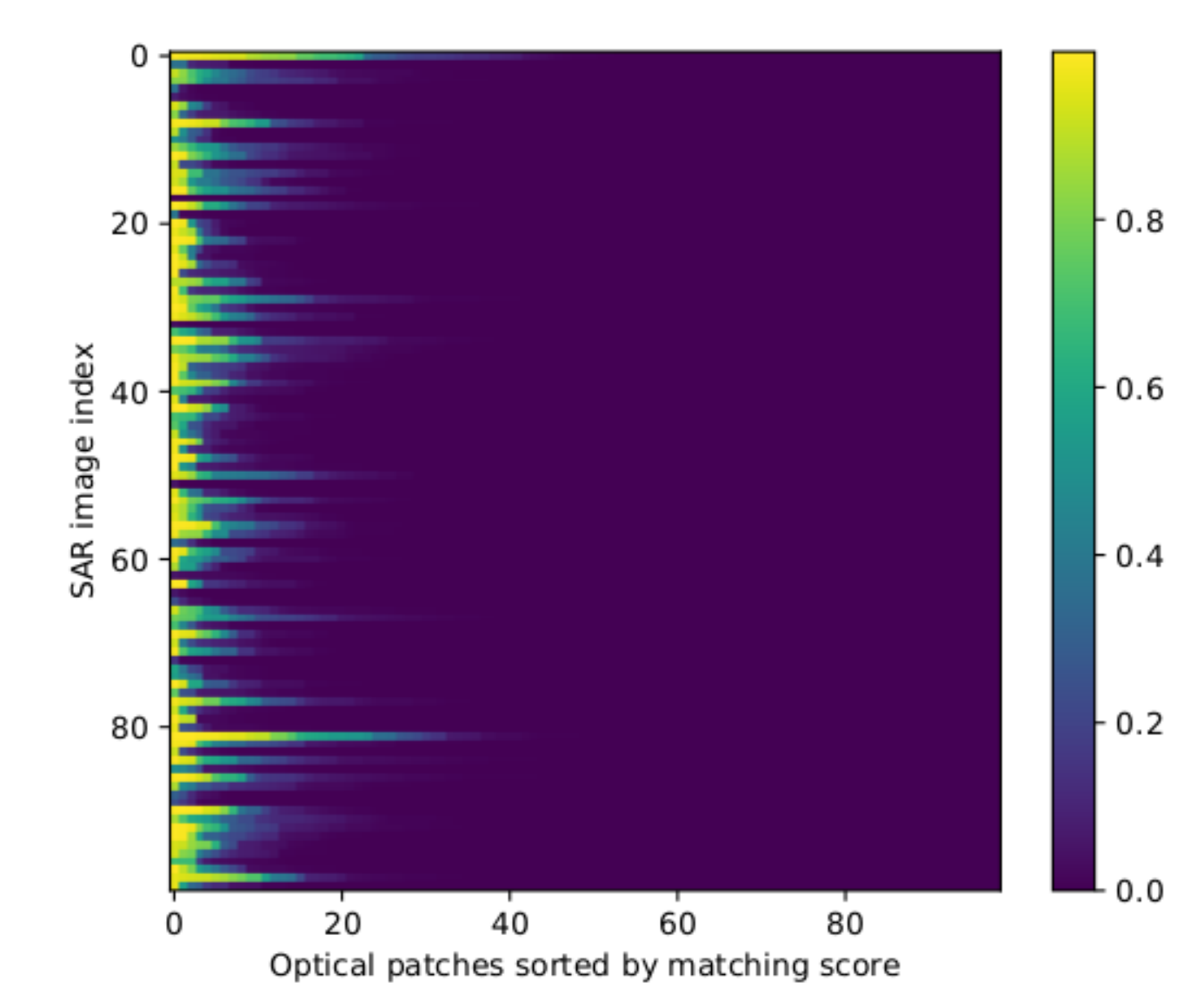}\\[-4pt]
\footnotesize a & \footnotesize b\\[2pt]
\end{tabular}
\caption{Results of key-point matching experiment. a) A confusion matrix showing the matching scores for all SAR and optical key-point patches. b) depicts the spread of incorrect matches, ordered by the similarity score.}\label{fig:fptmatch}
\end{figure}

In Fig. \ref{fig:fptmatch}a we can see a matrix depicting the similarity scores of the various pair comparisons, where corresponding SAR and optical patches are given the same index number. It should be noted that in determining a binary value for correspondence, a threshold is applied to these similarity scores. Figure \ref{fig:fptmatch}b depicts the sorted scores for all non-similar optical patches, making it easier to see the number and strength of incorrect matches in the patch pool.



\begin{table}[htbp]
  \centering
  \footnotesize
  \caption{Confusion matrix values for different patch sizes}
    \begin{tabular}{cccccc}
    \toprule
    & 64   & 76    & 88   & 100  & 112 \\
    \midrule
TP    & 46.6\% & 61.6\% & 66.0\% & 69.8\% & 82.2\% \\
TN    & 86.2\% & 88.0\% & 88.2\% & 86.0\% & 89.8\% \\
    \bottomrule
    \end{tabular}%
  \label{tab:confMatrix}%
\end{table}%

\begin{figure*}[tbh]
\centering
\begin{tabular}{cccc}
\includegraphics[width=.22\textwidth]{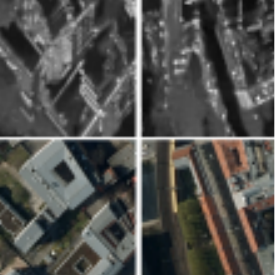} & \includegraphics[width=.22\textwidth]{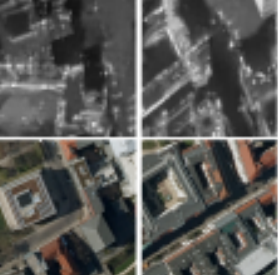} & \includegraphics[width=.22\textwidth]{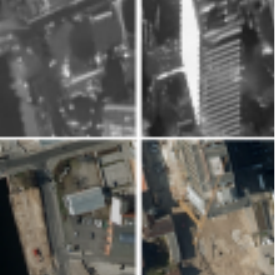} & \includegraphics[width=.22\textwidth]{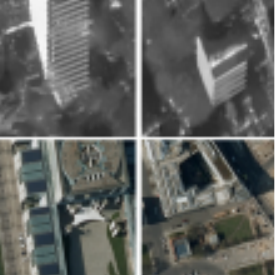}\\[-4pt]
\footnotesize True Positives & \footnotesize False Positives & \footnotesize False Negatives & \footnotesize True Negatives\\[2pt]
\end{tabular}
\caption{Exemplary patch correspondence results.}\label{fig:ColorResults}
\end{figure*}

\section{Discussion}
Generally, the results summarized in Section~\ref{sec:EvalResults} indicate a promising discriminative power of the proposed network. However, the following major points must be considered when interpreting the results.

\subsection{Influence of the Patch Size}
As Tab.~\ref{tab:confMatrix} and Fig.~\ref{fig:FPRplot} clearly indicate, the patch size strongly affects the discriminative power of the network. 
This result is likely due to the effects of distortions in SAR images, which are acquired in a range-based imaging geometry. Thus when patches are cropped to smaller regions we likely crop out defining features which are used for matching between the SAR and optical domain. This can be intuitively understood by referring to Fig.~\ref{fig:SAROPT} where we can see the effects of layover and multi-path reflections of the building in the SAR image, and a near top down view of the same building in the optical image. Taking away explanatory context will thus render the matching more difficult. All further discussion will be with reference to the results we obtained using the largest patch size, 112 pixels.

\subsection{Comments on the Discriminative Power of the Proposed Network}

In summary, our approach obtains an accuracy exceeding 77\% on a separate test dataset when fixing the false positive rate to 5\%, which falls into the same order of magnitude as what can be achieved using the powerful handcrafted HOPC descriptor in combination with an $L_2$-norm cost function \cite{Ye2017}. 

Furthermore, our approach produced a clear diagonal pattern in Fig. \ref{fig:fptmatch}a which depicts its ability to accurately determine the correct correspondence in a key-point matching scenario. Upon further investigation it was found that the network achieved 43\% top-1 matching accuracy and 74\% top-3 accuracy, while 8\% of points had no valid matches detected within the key-point set. This was found to be due to large amounts of layover and extreme differences in view point between the SAR and optical patches, see Fig. \ref{fig:ColorResults} False Negatives.


\subsection{Possible Reasons for False Predictions}
From the randomly chosen prediction examples displayed in Fig.~\ref{fig:ColorResults} it can be observed that many of the false positives and false negatives are erroneously matched under extreme differences in viewing angle between the SAR and optical patches. While this may partially be solvable by resorting to larger patch sizes, providing valuable context, there might be a need to exclude image parts with all too strong distortions from further processing.




\section{Conclusion}

In this paper, a pseudo-siamese CNN for learning to
identify corresponding patches in very high resolution SAR and optical images in a
fully automatic manner has been presented. A first evaluation has shown promising potential with respect to multi-sensor key-point matching procedures. In order to ensure transferability to other applications not based on key-points, e.g. dense matching, we will work on the generation of additional training patches, whose center pixel does not rely on specific key-points. In addition, we will test the approach on data coming from a completely different sources. In the end, we expect our work to help paving the way for generalized SAR-optical image matching procedures.


%

\appendices

\section*{Acknowledgment}
We thank the NVIDIA Corporation for donating the Titan X Pascal GPU used in this research.

\ifCLASSOPTIONcaptionsoff
  \newpage
\fi



%
\bibliographystyle{IEEEbib}
\bibliography{17GRSL}

\end{document}